\documentclass[twocolumn,aps,superscriptaddress,prb]{revtex4-1}
\usepackage{graphicx}
\usepackage{amssymb}
\usepackage{amsmath}
\usepackage{bm}
\usepackage{color}

\def\Journal #1,#2,#3,#4#5#6#7{#1 {\bf #2}, #3 (#4#5#6#7)}

\def\Vec{\mathbf}

\def\lsim{\, \lower -0.3ex \hbox{$<$} \kern -0.75em \lower 0.7ex \hbox{$\sim$} \,}
\def\gsim{\, \lower -0.3ex \hbox{$>$} \kern -0.75em \lower 0.7ex \hbox{$\sim$} \,}
\def\dg{$^\circ$}

\begin{document}

\title{Band structure and topological properties of twisted double bilayer graphenes\\
}



\author{Mikito Koshino}
\thanks{koshino@phys.sci.osaka-u.ac.jp}
\affiliation{Department of Physics, Osaka University,  Osaka 560-0043, Japan}

\begin{abstract}
We study the electronic band structure and the topological properties of
the twisted double bilayer graphene, 
or a pair of AB-stacked bilayer graphenes rotationally stacked on top of each other. 
We consider two different arrangements, AB-AB and AB-BA, which differ in the relative orientation.
For each system, we calculate the energy band and the valley Chern number using the continuum Hamiltonian.
We show that the AB-AB and the AB-BA have similar band structures, while the Chern numbers associated 
with the corresponding bands are completely different.
In the absence of the perpendicular electric field, in particular,
the AB-AB system is a trivial insulator when the Fermi energy is in a gap,
while the AB-BA is a valley Hall insulator. 
Also, the lowest electron and hole bands of the AB-AB 
are entangled by the symmetry protected band touching points,
while they are separated in the AB-BA.
In both cases, the perpendicular electric field immediately opens an energy gap at the charge neutral point, 
where the electron branch becomes much narrower than the hole branch, 
due to the significant electron-hole asymmetry.
\end{abstract}

\maketitle

\section{Introduction}

The electronic property in a stack of two-dimensional (2D) materials sensitively depends 
on the relative twist angle $\theta$ between the adjacent layers,
and we often have dramatic angle-dependent phenomena which are 
never observed in an isolated layer. 
The best known example is the twisted bilayer graphene (twisted BLG), or
a rotationally stacked pair of monolayer graphenes,
where a long-period moir\'{e} interference pattern significantly modifies the Dirac dispersion
\cite{lopes2007graphene,mele2010commensuration,trambly2010localization,shallcross2010electronic,morell2010flat,bistritzer2011moirepnas,moon2012energy,de2012numerical,moon2013opticalabsorption,weckbecker2016lowenergy}.
Recently, the superconductivity and correlated insulating 
states are discovered in the magic-angle twisted BLG with extremely flat bands,
\cite{cao2018unconventional,cao2018mott, yankowitz2019tuning}
and it is followed by a number of theoretical studies on the detailed properties of the flat bands
and the possible mechanism of the superconductivity.
\cite{yuan2018model,po2018origin,xu2018topological,kang2018symmetry,koshino2018maximally,ochi2018possible,isobe2018unconventional,dodaro2018phases,padhi2018doped,wu2018theory,tarnopolsky2019origin,zou2018band}
Graphene on hexagonal boron nitride (hBN) also exhibits 
the moir\'{e}-induced physics such as the formation of the secondary Dirac bands 
and the miniband structure. \cite{kindermann2012zero, wallbank2013generic, mucha2013heterostructures, jung2014ab, moon2014electronic,dean2013hofstadter,ponomarenko2013cloning,hunt2013massive,yu2014hierarchy}  
A recent experiment reported the correlated insulating states in ABC-trilayer graphene on hBN,
which is tunable by the external gate electric field. \cite{chen2019evidence}
Controlling the twist angle in a stack of 2D materials provides powerful means to manipulate quantum properties of the electronic systems.

\begin{figure}
\begin{center}
\leavevmode\includegraphics[width=0.81\hsize]{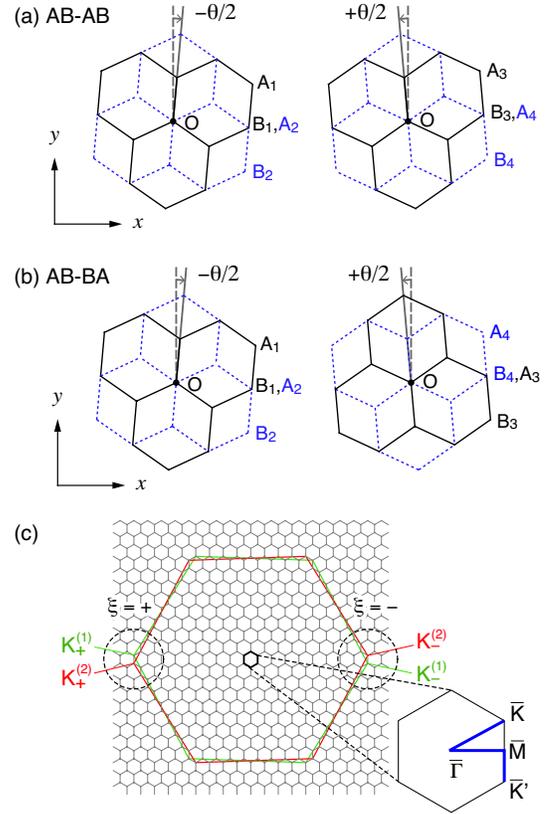}
\end{center}
\caption{
(a) Atomic structure of the twisted AB-AB double BLG
and (b) that of the twisted AB-BA double BLG.
(c) Brillouin zone folding in the double BLG.
Two large hexagons represent the first Brillouin zones of the first bilayer graphene, 
and the small hexagon is the moir\'{e} Brillouin zone.
}
\label{fig_lattice_BZ}
\end{figure}

In this paper, we study a different type of moir\'{e} system, the twisted double bilayer graphene, 
which is composed of a pair of AB-stacked BLGs rotationally stacked on top of each other. 
The AB-stacked BLG is the most stable form of bilayer graphene which has the stacking structure of graphite.
\cite{mccann2013electronic}
Here we consider two different arrangements, AB-AB and AB-BA, as illustrated 
in Figs.\ \ref{fig_lattice_BZ}(a) and 1(b), respectively,
where the AB-BA is obtained just by 180$^\circ$ rotation of the second BLG in the AB-AB.
For each case, we derive the continuum Hamiltonian by extending the approach for
the twisted BLG\cite{lopes2007graphene,bistritzer2011moirepnas,kindermann2011local,PhysRevB.86.155449,moon2013opticalabsorption,koshino2015interlayer,koshino2015electronic,weckbecker2016lowenergy}, 
and calculate the energy bands as well as the valley Chern numbers.
Here we include the interlayer asymmetric potential $\Delta$ induced by the gate electric field.

The energy band structures are found to be similar between the AB-AB and the AB-BA, 
but the topological nature is different.
In the absence of $\Delta$, the lowest electron and hole bands of the AB-AB 
are entangled by the symmetry protected band touching points,
while they are separated in the AB-BA due to the different space symmetry.
In both cases, the asymmetric potential $\Delta$ immediately opens an energy gap
at the charge neutral point.
We find that the graphite band parameters such as $\gamma_3$ and $\gamma_4$
play an important role in the electron-hole asymmetry, where the electron band
becomes much narrower than the hole band as increasing $\Delta$.

The crucial difference between AB-AB and AB-BA is found in the Chern number.
In the absence of $\Delta$, in particular,  the AB-AB double bilayer becomes a trivial insulator because 
the symmetry requires all the Chern numbers to vanish,
while the AB-BA is a valley Hall insulator with finite Chern number.
We demonstrate the evolution of the Chern numbers as a function of $\Delta$,
where we see that the energy bands of AB-AB and AB-BA  carry completely different topological numbers,
even though the band structures are similar.
The difference in the Chern number would be observed by the 
measurement of the valley Hall conductivity \cite{mak2014valley,shimazaki2015generation},
and also by the Landau level structure in the magnetic field.

This paper is organized as follows: In Sec.\ \ref{sec_atomic}, 
we define the lattice structures of AB-AB and AB-BA double bilayers,
and then introduce a continuum Hamiltonian for each system in Sec.\ \ref{sec_continuum}.
In Sec.\ \ref{sec_band}, we study the band structures and the evolution of Chern numbers
as a function of the twist angle and the asymmetric potential,
where we discuss in detail about similarity and difference between the two systems.
A brief conclusion is presented in Sec.\ \ref{sec_conclusion}.

\section{Atomic structure}
\label{sec_atomic}

The AB-stacked BLG is composed of a pair of monolayer graphenes,
with four atoms in the unit cell, labeled $A_1$, $B_1$ on the layer 1 (upper layer) 
and $A_2$, $B_2$ on the layer 2 (lower layer). \cite{mccann2013electronic}
The two graphene layers are arranged so that  $B_1$ and  $A_2$ are vertically located.
We refer to these two atomic sites as dimer sites because the electronic orbitals on them are strongly coupled. 
The other two atoms, $A_1$ and $B_2$ are directly above or below the hexagon center of the other layer,
and are referred to as non-dimer sites.

We compose the twisted AB-AB double bilayer graphene by stacking the first AB-stacked BLG 
(layers 1 and 2) on top of the second AB-stacked BLG (layers 3 and 4), as Fig.\ \ref{fig_lattice_BZ}(a).
We start from the non-rotated geometry where $B_1$, $A_2$, $B_3$ and  $A_4$ are
vertically aligned at the origin $O$, and then rotate the first and the second BLGs
around $O$ by $-\theta/2$ and $+\theta/2$, respectively.
The system has a three-fold in-plane rotation $C_{3z}$ symmetry along the $z$-axis (perpendicular to the layer), 
and a two-fold rotation $C_{2x}$ along the $x$-axis.
The twisted AB-BA double bilayer can be defined just by rotating the second BLG (layer 3 and 4)
of the AB-AB by 180$^\circ$ as in Fig.\ \ref{fig_lattice_BZ}(b),
where we flip the definition of $A$ site and $B$ site for layer 3 and 4.
The system is symmetric under a three-fold in-plane rotation $C_{3z}$ 
and a two-fold rotation $C_{2y}$ along the $y$-axis.

We define  $\Vec{a}_1 = a(1,0)$ and $\Vec{a}_2 = a(1/2,\sqrt{3}/2)$ 
as the lattice vectors of the initial BLGs before the rotation, where $a \approx 0.246\,\mathrm{nm}$ 
is the lattice constant of graphene. The corresponding reciprocal lattice vectors 
are $\Vec{a}^*_1 = (2\pi/a)(1,-1/\sqrt{3})$ and $\Vec{a}^*_2=(2\pi/a)(0,2/\sqrt{3})$.
After the rotation, the lattice vector of the $l$-th BLG is given by $\Vec{a}_i^{(l)} =R(\mp \theta/2)\Vec{a}_i$ 
with $\mp$ for $l=1,2$,  respectively, where $R(\theta)$ represents the rotation matrix by $\theta$.
The reciprocal lattice vectors become $\Vec{a}_i^{*(l)} =R(\mp \theta/2)\Vec{a}^*_i$. 
In a small $\theta$, the reciprocal lattice vectors for the moir\'{e} pattern is given by
$ \Vec{G}^{\rm M}_{i}  =  \textbf{a}^{*(1)}_i - \textbf{a}^{*(2)}_i \, (i=1,2)$,
and the real-space lattice vectors $\Vec{L}^{\rm M}_{j}$ can then be obtained from
$\Vec{G}^{\rm M}_i\cdot\Vec{L}^{\rm M}_{j} = 2\pi\delta_{ij}$.
A moir\'{e} unit cell is spanned by $\Vec{L}^{\rm M}_{1}$ and $\Vec{L}^{\rm M}_2$.
The lattice constant $L_{\rm M} = | \Vec{L}^{\rm M}_{1}|=| \Vec{L}^{\rm M}_2|$
is $L_{\rm M} =  a/[2\sin (\theta/2)]$.
Figure \ref{fig_lattice_BZ}(c) illustrates the Brillouin zone folding, 
where two large hexagons represent the first Brillouin zones of the first and the second BLGs, 
and the small hexagon is the moir\'{e} Brillouin zone of the twisted double BLG.
The graphene's Dirac points (the band touching points) 
are located  at $\Vec{K}^{(l)}_\xi = -\xi [2\Vec{a}^{(l)*}_1+\Vec{a}^{(l)*}_2]/3$ 
for the $l$-th BLG, where $\xi=\pm 1$ is the valley index.
We label the symmetric points of the moir\'{e}  Brillouin zone as
$\bar{\Gamma}$, $\bar{M}$, $\bar{K}$ and $\bar{K'}$ as in Fig.\ \ref{fig_lattice_BZ}(c).

\section{Continuum Hamiltonian}
\label{sec_continuum}


 To describe the electronic band structure of the twisted double bilayers,
we adopt the continuum method based on the Dirac Hamiltonian.\cite{lopes2007graphene,bistritzer2011moirepnas,kindermann2011local,PhysRevB.86.155449,moon2013opticalabsorption,koshino2015interlayer,koshino2015electronic,weckbecker2016lowenergy}
The validity of the continuum model was verified for twisted BLG 
by the direct comparison to the tight-binding model. \cite{moon2013opticalabsorption,weckbecker2016lowenergy}
We define the Bloch bases of $p_z$ orbitals at sublattice
 $X= A_1,B_1, \cdots ,A_4,B_4$ as
 $ |\Vec{k},X\rangle = N^{-1/2} \sum_{\Vec{R}_{X}} e^{i\Vec{k}\cdot\Vec{R}_{X}}$,
where $|\Vec{R}_{X} \rangle$ is the atomic $p_z$ orbital at the site $\Vec{R}_{X}$, 
$\Vec{k}$ is the two-dimensional Bloch wave vector 
and $N$  is the number of same sublattices in the system. 
The continuum Hamiltonian for 
 twisted AB-AB double bilayer graphene at small twist angle $\theta (\ll 1)$
is written in 8 $\times $ 8 matrix for the Bloch bases of 
 $(A_1,B_1,A_2,B_2,A_3,B_3,A_4,B_4)$ as
 \begin{align}
&	
{H}_{\textrm{AB-AB}} = 
	\begin{pmatrix}
		H_0(\Vec{k}_1) & g^\dagger(\Vec{k}_1) & & \\
		g(\Vec{k}_1) & H'_0(\Vec{k}_1) & U^\dagger  &\\
		& U & H_0(\Vec{k}_2) & g^\dagger(\Vec{k}_2)  \\
		& & g(\Vec{k}_2) & H'_0(\Vec{k}_2) \\
	\end{pmatrix}  + V,
	\label{eq_AB-AB}
\end{align}
where $ \Vec{k}_l = R(\pm \theta/2)({\Vec{k}}-\Vec{K}^{(l)}_\xi)$
with $\pm$ for $l=1$ and 2, respectively, and 
\begin{align}
& H_0(\Vec{k}) 
=
\begin{pmatrix}
0  & -\hbar v k_- \\
-\hbar v k_+ & \Delta'
\end{pmatrix},
\,
H'_0(\Vec{k}) 
=
\begin{pmatrix}
\Delta'  & -\hbar v k_- \\
-\hbar v k_+ & 0
\end{pmatrix} 
\\ 
& g(\Vec{k}) 
=
\begin{pmatrix}
\hbar v_4 k_+  & \gamma_1 \\
\hbar v_3 k_-  & \hbar v_4 k_+
\end{pmatrix},
\end{align}
with $k_\pm = \xi k_x \pm i k_y$.
$H_0$ and $H'_0$ are the Hamiltonian of monolayer graphene
where $\Delta' = 0.050 $eV\cite{mccann2013electronic} 
represents the on-site potential of dimer sites with respect to non-dimer sites.
The parameter $v$ is the band velocity of monolayer graphene,
and it is taken as $\hbar v /a = 2.1354$ eV.\cite{moon2013opticalabsorption,koshino2018maximally}
The matrix $g$ describes the interlayer coupling of the AB-stacked BLG,
where $\gamma_1 = 0.4$eV is the coupling between dimer sites, and 
$v_3$ and $v_4$ are related to diagonal hoppings  $\gamma_3 = 0.32$eV and $\gamma_4=0.044$ eV
with the relation $v_i = (\sqrt{3}/2) \gamma_i a /\hbar \, (i=3,4)$. \cite{mccann2013electronic} 
In the AB-stacked BLG,  $v_3$ is responsible for the trigonal warping of the energy band
and $v_4$ is for the electron-hole asymmetry.

The matrix $U$ is the moir\'{e} interlayer coupling between twisted layers given by
\cite{bistritzer2011moirepnas,moon2013opticalabsorption,koshino2018maximally}
\begin{align}
 U &= 
\begin{pmatrix}
u & u'
\\
u' & u
\end{pmatrix}
+
\begin{pmatrix}
u & u'\omega^{-\xi}
\\
u'\omega^\xi & u
\end{pmatrix}
e^{i\xi \Vec{G}^{\rm M}_1\cdot\Vec{r}}
\nonumber\\
& 
\qquad \qquad \qquad +
\begin{pmatrix}
u & u'\omega^\xi
\\
u'\omega^{-\xi} & u
\end{pmatrix}
e^{i\xi(\Vec{G}^{\rm M}_1+\Vec{G}^{\rm M}_2)\cdot\Vec{r}},
\label{eq_interlayer_matrix}
\end{align}
where $\omega =e^{2\pi i/3}$,
$u = 0.0797$eV and $u' = 0.0975$eV \cite{koshino2018maximally}
are the amplitudes of diagonal and off-diagonal terms, respectively.
The difference between $u$ and $u'$ effectively describe the out-of-plane corrugation effect,
which enhances the energy gaps between the lowest energy bands 
and the excited bands.\cite{koshino2018maximally,nam2017lattice,tarnopolsky2019origin}
Lastly,  $V$ is the interlayer asymmetric potential,
\begin{align}
& V = 
 \begin{pmatrix} 
 \frac{3}{2}\Delta \hat{1} &&& \\
&  \frac{1}{2}\Delta \hat{1} && \\
&&   -\frac{1}{2}\Delta \hat{1} & \\
&&&  -\frac{3}{2}\Delta \hat{1} 
 \end{pmatrix},
\end{align}
where $\hat{1}$ is $2\times 2$ unit matrix, and $\Delta$ represents 
the difference in the electrostatic energy between the adjacent layers.
Here we simply assumed the perpendicular electric field is constant.

Noting that the lattice structure of the AB-AB double bilayer has $C_{2x}$ symmetry
and also the valley degree of freedom $\xi=\pm$ is unchanged under $C_{2x}$,
the Hamiltonian $H_\textrm{AB-AB}$ of each single valley commutes with $C_{2x}$,
given that the asymmetric potential $\Delta$ is absent.

Similarly, the Hamiltonian of the twisted AB-BA double bilayer graphene is given by 
 \begin{align}
&	
{H}_{\textrm{AB-BA}} = 
\begin{pmatrix}
H_0(\Vec{k}_1) & g^\dagger(\Vec{k}_1) & & \\
g(\Vec{k}_1) & H'_0(\Vec{k}_1) & U^\dagger  &\\
& U & H'_0(\Vec{k}_2) & g(\Vec{k}_2)  \\
& & g^\dagger(\Vec{k}_2) & H_0(\Vec{k}_2) \\
\end{pmatrix} 
 + V.
\label{eq_AB-BA}
\end{align}
where $H_0(\Vec{k}_2)$ and $H'_0(\Vec{k}_2)$ are interchanged and 
also $g(\Vec{k}_2)$ and $g^\dagger(\Vec{k}_2)$ are swapped
in ${H}_{\textrm{AB-AB}}$.
The lattice structure of the AB-BA double bilayer has $C_{2y}$ symmetry,
and $C_{2y}$ interchanges the valleys $\xi=\pm$.
As a result, the Hamiltonian $H_\textrm{AB-BA}$ with $\Delta = 0$
commutes with $C_{2y} T$, where $T$ is the time reversal operator.

 The calculation of the energy bands and the eigenstates 
 is performed in the $k$-space picture.
For a single Bloch vector $\Vec{k}$ in the moir\'{e} Brillouin zone,   
the interlayer coupling $U$  hybridizes the graphene's eigenstates at
$\Vec{q} = \Vec{k} + \Vec{G}$, where $\Vec{G} = m_1  \Vec{G}^{\rm M}_1 + m_2  \Vec{G}^{\rm M}_2$
and $m_1$ and $m_2$ are integers.
The low-energy eigenstates can be obtained by numerically diagonalizing the Hamiltonian
within the limited number of $\Vec{q}$'s inside the cut-off circle $|\Vec{q}-\Vec{q}_0| < q_c$.
Here $\Vec{q}_0$ is taken as the midpoint between $\Vec{K}^{(1)}_\xi$ and $\Vec{K}^{(2)}_\xi$,
and $q_c$ is set to $4 |\Vec{G}^{\rm M}_1|$.
The calculation is done independently for each of $\xi=\pm$ as the intervalley coupling can be neglected
in small twist angles.

We calculate the Chern number of moir\'{e} subbands by the standard definition,
\begin{align}
& C_n = \frac{1}{2\pi}\int_{\rm MBZ} {\cal F}_{n,\Vec{k}}  \, d^2k,
\end{align}
where $n$ is the band index, 
MBZ represents the moir\'{e} Brillouin zone, and $F_{n,\Vec{k}}$ is the Berry curvature defined by
\begin{align}
{\cal F}_{n,\Vec{k}}  = \frac{\partial a^{(y)}_{n,\Vec{k}}}{\partial k_x}  -  \frac{\partial a^{(x)}_{n,\Vec{k}}}{\partial k_y}, \quad 
a^{(\mu)}_{n,\Vec{k}} = \frac{1}{i} \langle u_{n,\Vec{k}} |  \frac{\partial}{\partial k_\mu} | u_{n,\Vec{k}} \rangle,
\end{align}
where $u_{n,\Vec{k}}$ is the Bloch wave function of $n$-th subband.
We numerically calculate the Chern numbers using the discretizing method. \cite{fukui2005chern}

The symmetry imposes constraints on the Chern number.
For the AB-AB double bilayer at $\Delta =0$, the $C_{2x}$ 
symmetry requires ${\cal F}_{n,(k_x,-k_y)} = - {\cal F}_{n,(k_x,k_y)}$,
so that the Chern number of each single band must vanish.
In the AB-BA double bilayer at $\Delta =0$,  the $C_{2y} T$ 
symmetry requires ${\cal F}_{n,(k_x,-k_y)} = {\cal F}_{n,(k_x,k_y)}$, and the Chern number can be finite. 

\begin{figure}
\begin{center}
\leavevmode\includegraphics[width=0.95\hsize]{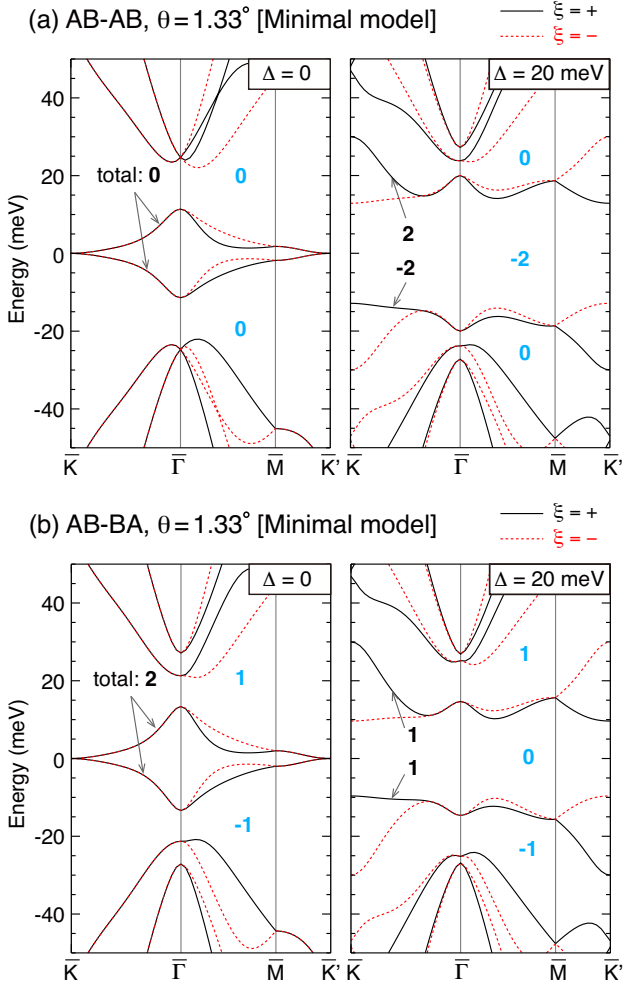}
\end{center}
\caption{(a) Band structure of the twisted AB-AB double bilayer at the twist angle $\theta= 1.33^\circ$
with $\Delta = 0$ and 20 meV, calculated by the minimal model.
(b) Corresponding plots for the twisted AB-BA double bilayer.
Black numbers indicate the Chern numbers for the energy bands in $\xi =+$,
and the blue numbers between the bands are the integrated Chern numbers 
summed over all the energy bands of $\xi =+$ below.
The Chern numbers for $\xi =-$ bands are opposite in sign.
}
\label{fig_band_1.33minimal}
\end{figure}

\section{Band structures and topological properties}
\label{sec_band}

\subsection{Minimal model}
\label{sec_minimal}

Before calculating the band structure with all the band parameters fully included,
it is intuitive to consider the minimal model which neglects the relatively small parameters, $v_3$, $v_4$, $\Delta'$ 
and the rotation matrix $R(\pm \theta/2)$ in the definitions of $\Vec{k}^{(1)}$ and $\Vec{k}^{(2)}$.
Then the AB-AB Hamiltonian Eq.\ (\ref{eq_AB-AB}) has a fictitious particle-hole symmetry
similar to TBG\cite{moon2013opticalabsorption},
\begin{align}
\Sigma^{-1} H_{\textrm{AB-AB}} \Sigma = - H^*_{\textrm{AB-AB}}, \nonumber\\
\Sigma = 
\begin{pmatrix}
&&& \sigma_x \\
&& -\sigma_x &\\
& \sigma_x &&\\
- \sigma_x  &&&
\end{pmatrix}.
\label{eq_e-h_sym_AB-AB}
\end{align}
This immediately leads to the electron-hole symmetry in the energy bands,
$E_{n,\Vec{k}} = -E_{-n,-\Vec{k}}$, and also the anti-symmetric relation in the Chern number,
$C_{-n} = - C_n$, where $n$ and $-n$ stand for the band indexes of
the corresponding electron and hole bands, respectively.
Note that Eq.\ (\ref{eq_e-h_sym_AB-AB}) holds even 
in the presence of the interlayer asymmetric potential $\Delta$.

The AB-BA Hamiltonian Eq.\ (\ref{eq_AB-BA}) has a different type of symmetry
between the electron and the hole bands,
\begin{align}(
\Sigma'^{-1} \tilde{P}) H_{\textrm{AB-BA}} (\tilde{P} \Sigma') = - H_{\textrm{AB-BA}}, \nonumber\\
\Sigma' = 
\begin{pmatrix}
&&& \hat{1} \\
&& - \hat{1}  &\\
&  \hat{1}  &&\\
- \hat{1} &&&
\end{pmatrix},
\label{eq_e-h_sym_AB-BA}
\end{align}
where $\tilde{P}$ is a space inversion operator which works on envelop function as
$\tilde{P} F_X(\Vec{r}) = F_X(-\Vec{r})$,
while it does not change the sublattice degree of freedom ($X=A_1,B_1,\cdots$).
This again forces the electron-hole symmetry $E_{n,\Vec{k}} = -E_{-n,-\Vec{k}}$,
but the Chern number becomes electron-hole symmetric, $C_{-n} = C_n$,
because the operation lacks the complex conjugate.

Figures \ref{fig_band_1.33minimal}(a) and (b) show the minimal-model band structure,
calculated for the AB-AB double bilayer and the AB-BA double bilayer, respectively, 
at the twist angle $\theta= 1.33^\circ$ with $\Delta = 0$ and 20 meV. 
The band structures of the two systems closely resemble each other.
At $\Delta =0$, we have a pair of energy bands touching at Dirac point,
which are isolated from the excited bands by energy gaps as in the twisted BLG. \cite{koshino2018maximally}
A finite $\Delta$ immediately opens an energy gap at the charge neutral point.
This is in a sharp contrast to the twisted BLG, where the perpendicular electric field never opens a gap 
at the charge neutral point, because the band touching is protected by $C_2 T$ symmetry.
Now the twisted double bilayer lacks $C_2$ symmetry.

Although the band structures are pretty much similar between the AB-AB and the AB-BA cases,
the properties of the Chern number are completely different.
In Fig.\ \ref{fig_band_1.33minimal}, the black numbers indicate the Chern numbers
of the central two bands in $\xi =+$ valley,
and the blue numbers between the bands are the integrated Chern numbers 
summed over all the energy bands of $\xi =+$ below.
Because of the time reversal symmetry, the Chern number of $\xi =-$ valley is opposite in sign to $\xi =+$. 
We actually see the expected relation $C_{-n} = - C_n$ for the AB-AB, and $C_{-n} = C_n$ for the AB-BA.
In the absence of the asymmetric potential $\Delta$, the Chern numbers all vanish in the AB-AB because of the 
rigorous symmetry $C_{2x}$ mentioned in the previous section, 
while it is finite in the AB-BA.
When the Fermi energy is inside one of those gaps, therefore,
the AB-BA double bilayer is a valley Hall insulator, while AB-AB is a trivial insulator.
The Chern numbers can be finite in the AB-AB  once the asymmetric potential $\Delta$ is switched on,
because it breaks $C_{2x}$.
The integrated Chern numbers inside the central gap is $-2$ in the AB-AB, while 0 in the AB-BA. 
This is just equal to the sum of Chern numbers of two independent gapped BLGs,
which is $-1$ for the AB stack while $+1$ for the BA stack. \cite{martin2008topological,koshino2008electron,zhang2013valley,vaezi2013topological}

\begin{figure*}
\begin{center}
\leavevmode\includegraphics[width=0.75\hsize]{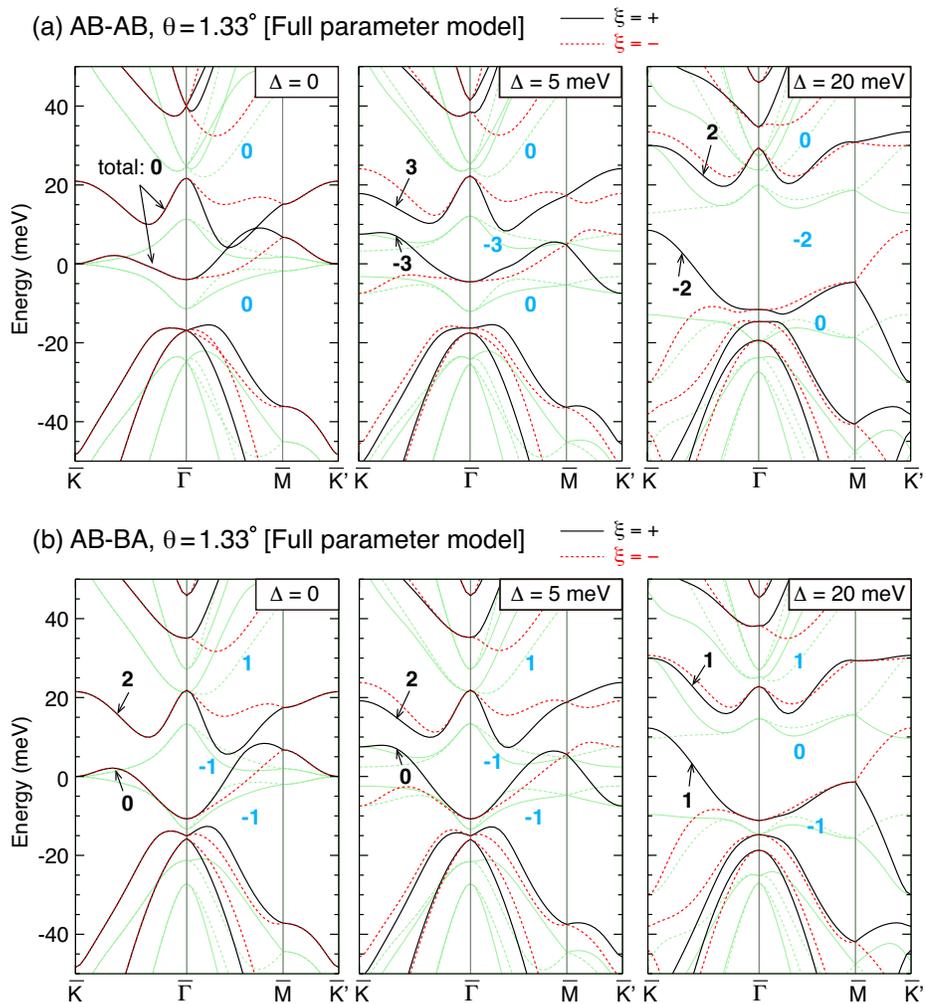}
\end{center}
\caption{(a) Band structure of the AB-AB double bilayer at the twist angle $\theta= 1.33^\circ$
with $\Delta = 0, 5$ and 20 meV, calculated by the full parameter model,
(b) Corresponding plots for the AB-BA double bilayer.
Thin green lines indicate the energy bands of the minimal model [Fig.\ \ref{fig_band_1.33minimal}].
}
\label{fig_band_1.33full}
\end{figure*}

\begin{figure*}
\begin{center}
\leavevmode\includegraphics[width=1.\hsize]{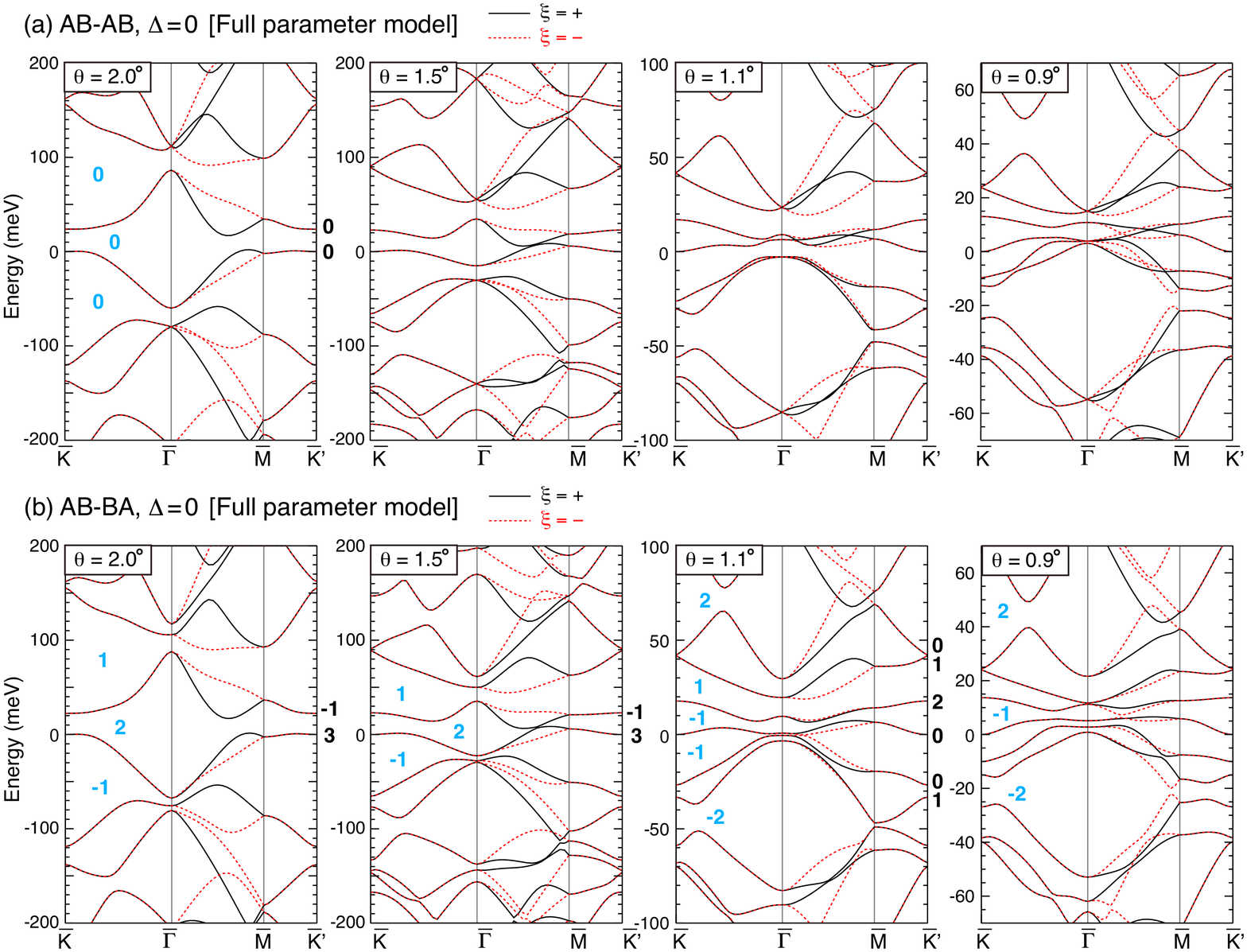}
\end{center}
\caption{
(a) Band structure of the AB-AB double bilayer at various twist angles with $\Delta = 0$, 
calculated by the full parameter model.
(b) Corresponding plots for the AB-BA double bilayer.
}
\label{fig_band_angle_dep}
\end{figure*}

\subsection{Full parameter model}

Inclusion of the additional band parameters neglected in the minimal model
causes a significant change particularly in the low-energy sector.
Figures \ref{fig_band_1.33full}(a) and (b) show the full-parameter band structure
of the AB-AB and the AB-BA double bilayers, respectively, at the twist angle $\theta= 1.33^\circ$
with $\Delta = 0, 5$ and 20 meV.
The thin green lines indicate the energy bands of the minimal model [Fig.\ \ref{fig_band_1.33minimal}]
for quantitative comparison.
We see that the energy bands are now electron-hole asymmetric 
because the fictitious symmetry of Eq.\ (\ref{eq_e-h_sym_AB-AB}) or Eq.\ (\ref{eq_e-h_sym_AB-BA})
is broken.
The band structures of AB-AB and AB-BA are still similar,
but there are several important differences.
At $\Delta =0$, in particular,
the central energy bands of the AB-AB are touching at two points on the $\bar{\Gamma}-\bar{M}$ line,
while they are anti-crossing in the AB-BA.
The band touching of the former is protected by the $C_{2x}$ symmetry.
Since the $k$-points on $\bar{\Gamma}-\bar{M}$ are invariant under $C_{2x}$ operation,
the Bloch states on the line can be characterized by the eigenvalues of $C_{2x}$.
The energy bands crossing at the center actually have
the opposite eignevalues $C_{2x} = \pm 1$, so that they are never hybridized. 
The energy bands form a two-dimensional Dirac cone around each band touching point.
Because of $C_{3}$ symmetry, we have six touching points in each single valley.
Note that the energy bands of $\xi=-$ is just 180$^\circ$ rotation of $\xi=+$ band,
so the band touching of $\xi=-$ are not seen in the figure.

In increasing $\Delta$, we see that 
the upper central band (the first conduction band) becomes much narrower than
the lower central band (the first valence band), in both of the AB-AB and the AB-BA.
As a result, the energy gap just above the upper band survives
in relatively large $\Delta$,
while the gap below the lower band is easily masked by the wide dispersion.

The properties of the Chern number are mostly carried over from the minimal model.
A difference is seen in $\Delta = 5$ meV in the AB-AB case, where the central bands have the 
Chern number $\pm 3$, unlike $\pm 2$ in the minimal model.
This is attributed to the six Dirac points at $\Delta =0$,
each of which contributes to the Berry curvature $\pi$ when gapped out.
In even increasing $\Delta$, we have a band touching at $\bar{K}$ around $\Delta \sim 9$ meV,
where the Chern number $+1$ is transferred from the lower central band to the higher central band.
As a result, the Chern number of the central bands becomes $\pm 2$ as in the minimal model.
A similar topological change is also observed in the AB-BA double bilayer,
where the Chern numbers of the central two bands change 
from $(2,0)$ to $(1,1)$.

Finally, we present in Fig.\ \ref{fig_band_angle_dep} the twist angle dependence of the band structure in 
(a) the AB-AB and (b) the AB-BA double bilayers with $\Delta = 0$.
For the AB-AB, the central electron and hole bands are separated by an energy gap at $\theta = 2^\circ$,
and they get closer in decreasing $\theta$.
At $\theta \sim 1.44^\circ$, there is a quadratic band touching on $\bar{\Gamma}-\bar{M}$ line,
and a pair of the Dirac points are formed below that angle.
Those band touching points are protected by $C_{2x}$ symmetry as already argued.
In even lower angles, the central energy bands become narrower and narrower,
and at the same time the higher excited bands collides with the central bands.
At $\theta = 1.1^\circ$ and $0.9^\circ$, we have an insulating gap between 
the third and the fourth valence bands. 
Because of $C_{2x}$ symmetry, the Chern number is zero everywhere as long as $\Delta=0$.

In the AB-BA case, we have a similar evolution of the band structure, 
while the Chern number is generally non-zero.
At $\theta = 2^\circ$, the charge neutral point is a valley Hall insulator with the Chern number 2.
In decreasing $\theta$, we have a topological change at $\theta \sim 1.45^\circ$,
where the Chern number 3 is transferred from the lower band to the higher band
through the three touching points arranged in 120\dg symmetry.
Unlike the AB-AB, the band touching occurs only at the topological transition,
and the bands are separated again after the transition.
In smaller angles less than 1\dg, the central bands touch with the excited bands,
and there are complex topological changes between them.
We have a new insulating gap between the third and the fourth valence bands, where the  Chern number is $-2$.

\begin{figure*}
\begin{center}
\leavevmode\includegraphics[width=0.75\hsize]{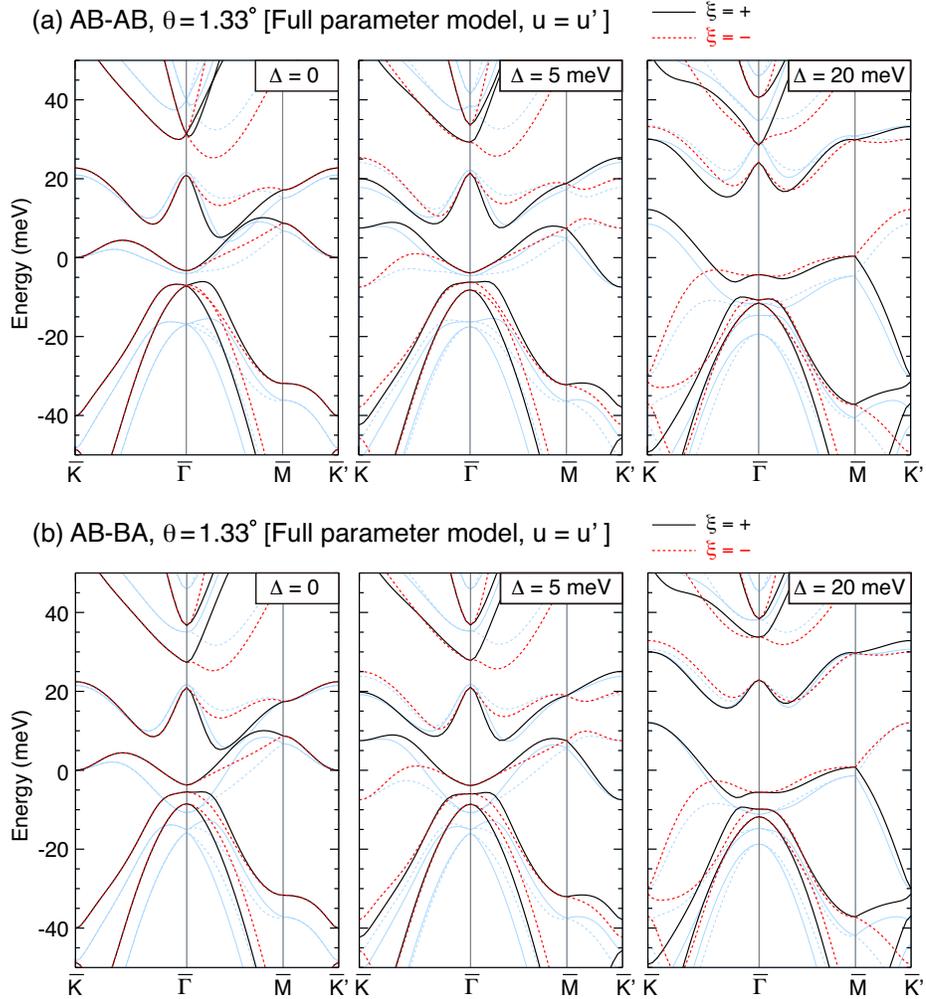}
\end{center}
\caption{(a) Band structures of the AB-AB double bilayer at $\theta= 1.33^\circ$
calculated by the full parameter model with $u=u' =  0.0975$eV.
Thin blue lines are the original results with $u = 0.0797$eV and $u' = 0.0975$eV [Fig.\ \ref{fig_band_1.33full}].
(b) Corresponding plots for the AB-BA double bilayer.
}
\label{fig_band_1.33full_no_corr}
\end{figure*}

In this work, we assumed different parameters $u$ and $u'$
to describe possible corrugation effect, where the adopted values, $u = 0.0797$eV and $u' = 0.0975$eV,
are taken from the twisted BLG (monolayer-monolayer) \cite{koshino2018maximally}.
In the twisted double BLG, however, the corrugation would be reduced to some extent
considering that bilayer graphene is stiffer than monolayer graphene,
and then the difference between $u$ and $u'$ should also decrease accordingly.
To see this effect, we present the band structure assuming $u=u' =  0.0975$eV
in Fig.\ \ref{fig_band_1.33full_no_corr}, where thin blue lines represent the original calculation
from Fig.\ \ref{fig_band_1.33full}.
We see that the qualitative features are similar, while 
the energy gaps between the lowest bands and the excited bands are smaller
in the $u = u'$ model
than in the $u \neq u'$ model, as in the twisted BLG. \cite{koshino2018maximally, nam2017lattice,tarnopolsky2019origin}.
The real situation should be somewhere between the two cases.

\section{Conclusion}
\label{sec_conclusion}
 We have studied the electronic band structure and the Chern numbers of
 AB-AB and AB-BA twisted double bilayer graphenes,
and found that the two systems have similar band structures, but with completely different 
topological properties. In the absence of the asymmetric potential $\Delta$ (perpendicular electric field), 
in particular,
the AB-BA double bilayer is a valley Hall insulator when the Fermi energy is in a gap,
while the AB-AB is a trivial insulator due to the symmetry constraint.
Also, the energy bands of the AB-AB in $\Delta= 0$ 
are entangled by the symmetry protected band touching points, while they are all separated in the AB-BA.
The common features shared by the two systems
is that a pair of narrow bands at the charge neutral point
are immediately gapped by applying the perpendicular electric field, unlike the twisted BLG (monolayer-monolayer).
There the graphite band parameters such as $\gamma_3$, $\gamma_4$ play an important role
in the electron-hole asymmetry, where the electron branch
becomes much narrower than the hole branch in increasing the perpendicular electric field.

\section*{Acknowledgments}
MK thanks the fruitful discussions with Jeil Jung, Pablo Jarillo-Herrero, 
Philip Kim, Eslam Khalaf, Jong Yeon Lee and Ashvin Vishwanath. 
MK acknowledges the financial support of JSPS KAKENHI Grant Number JP17K05496. 

{\it Note added:}
The recent preprints reported experimental observations of 
superconductivity and correlated insulating states in the twisted double bilayer graphenes.
\cite{shen2019observation,liu2019spin,cao2019electric}
During the completion of this work, 
we became aware of recent theoretical works on the electronic properties of twisted AB-AB double BLG \cite{chebrolu2019flatbands,choi2019intrinsic,lee2019theory}.
Just after the submission of this manuscript, we have come to notice a recent theoretical study on the 
electronic and topological properties on twisted multilayer graphene systems
with various stacking configurations. \cite{liu2019quantum}

\bibliography{twisted_BLG-BLG}

\end{document}